\documentclass{article}

\usepackage{vmargin,amssymb,amsmath,epsfig,latexsym,booktabs,braket,
gensymb,amsbsy,stackrel,slashed}

\newcommand{\tr}{\mathrm{Tr}}
\newcommand{\cN}{{\cal N}}
\newcommand{\cL}{{\cal L}}
\newcommand{\nFour}{$\cN \, = \, 4$ }
\newcommand{\rar}{\, \rightarrow \,}
\newcommand{\cO}{{\cal O}}
\newcommand{\eqsp}{\, = \,}
\newcommand{\la}{\langle}
\newcommand{\ra}{\rangle}

\newcommand{\beq}{\begin{equation}}
\newcommand{\eeq}{\end{equation}}

\newcommand{\phm}{\phantom{-}}

\begin{document}

\thispagestyle{empty}

\begin{flushright} HU-EP-23/05 \end{flushright}

\vskip 3.5 cm

\begin{center}

{\huge \textbf{Double excitations in the AdS(5)/CFT(4) \\[3 mm] integrable system
and the Lagrange operator}}

\vskip 1.5 cm

{\large Burkhard~Eden$^a$, Dennis~le~Plat$^a$, Anne~Spiering$^b$}
\vskip 1 cm

$^a$ Institut f\"ur Mathematik und Physik, Humboldt-Universit\"at zu Berlin, \\ Zum gro{\ss}en Windkanal 2, 12489 Berlin, Germany \\[2 mm]
$^b$ Niels Bohr Institute, University of Copenhagen,\\ 
Blegdamsvej 17, 2100 Copenhagen 0, Denmark

\vskip 1 cm

e-mail: eden@math.hu-berlin.de, diplat@physik.hu-berlin.de, anne.spiering@nbi.ku.dk

\end{center}

\vskip 2.5 cm

\textbf{Abstract:} It is argued that the integrable model for the planar spectrum of the AdS/CFT correspondence can accommodate for the full spectrum of excitations $D^{\alpha \dot \alpha}, \phi^{[IJ]}, \psi^I, \bar \psi_I, F^{\alpha \beta}, \tilde F^{\dot \alpha \dot \beta}$ (with $I,J \in 1 \ldots 4$) if double excitations are allowed for all three raising operators of the internal $SU(4)$ symmetry. We present a tree-level analysis of related creation amplitudes in the nested Bethe ansatz as well as in the original level-1 picture in which excitations of various flavours scatter by a true $S$-matrix. In the latter case, the creation amplitudes for all double excitations we encounter take a perfectly universal form. 

Building on these ideas we work out Bethe solutions and states relevant in the mixing problem concerning the on-shell Lagrangian of \nFour super Yang-Mills theory. Owing to the very existence of double excitations, the chiral Yang-Mills field strength tensor can be represented by the four fermions $\{\psi^{31}, \psi^{32}, \psi^{41}, \psi^{42}\}$ moving on a spin chain of length two. 
Our analysis remains restricted to leading order in the coupling, where the conformal eigenstate corresponding to the on-shell Lagrangian only comprises the pure Yang-Mills action. It should eventually be possible to augment our analysis to higher loop orders by incorporating coupling corrections in the relevant ingredients from the Bethe ansatz.
 
Finally, it was recently realised how structure constants for operators containing the hitherto hidden half of the excitations can be computed by the hexagon formalism. We use this for a first test of our conjecture for the on-shell Lagrangian, namely that its three-point function with two half-BPS operators of equal length ought to vanish.

\newpage

\section{Introduction and review}

When the BMN operator mixing problem \cite{bmn} was first connected \cite{Minaza} to an integrable system, it was in the so-called $so(6)$ sector of the \nFour super Yang-Mills theory in four dimensions. This sector comprises the six real scalar fields $\phi^m$, or equivalently complex linear combinations satisfying a reality constraint
\beq
\phi^{[IJ]} \, , \quad I,J \, \in \, \{1,2,3,4\} \, , \qquad \phi_{IJ} \eqsp \left(\phi^{IJ} \right)^* \eqsp \frac{1}{2} \epsilon_{IJKL} \phi^{KL} \, .
\eeq
The nested Bethe ansatz for $so(6)$ starts out like Bethe's ansatz for the Heisenberg spin chain (an $su(2)$ sector) in which, say, some fields $X \eqsp \phi^{24}$ move over a chain of vacuum sites $Z \eqsp \phi^{34}$. Equivalently, one may think of lowering operators $R_3^2$ moving over the vacuum. Each \emph{excitation}, or \emph{magnon}, $X$ is equipped with a quasi-momentum $p$ or the associated \emph{Bethe rapidity} $u$:
\beq
e^{i p} \, = \, \frac{u+\frac{i}{2}}{u-\frac{i}{2}} \, , \qquad u \eqsp \frac{1}{2} \cot\left(\frac{p}{2}\right)~. \label{momFac}
\eeq
When an excitation with rapidity $u_1$ overtakes another with rapidity $u_2$, a scattering phase
\beq
S^{11}(u_1,u_2) \, = \, \frac{u_1-u_2-i}{u_1-u_2+i} \label{S11}
\eeq
arises. 

On the $X$'s (so only the sites where $R_3^2$ has already acted) we can further lower by $R_2^1$ to $Y \eqsp \phi^{14}$ or by $R_4^3$ to $\bar Y \eqsp \phi^{23}$. For these excitations we introduce \emph{secondary} or \emph{auxilliary rapidities} $v,w$, respectively.
These two operations $R_2^1, R_4^3$ commute and they generate a left- and right- wing algebra, respectively. Unlike in the underlying $su(2)$ sector, or \emph{level 1}, of the problem, the level 2 and 3 excitations come with a special amplitude 
\beq
f^{21}(v,u) \eqsp \frac{1}{v-u-\frac{i}{2}} \, , \qquad f^{31}(w,v) \eqsp \frac{1}{w-u-\frac{i}{2}}
\eeq
when created from an underlying $X$ with rapidity $u$, and for moving them over some other $X(u')$ there are scattering matrices
\beq
S^{21}(v,u') \eqsp \, \frac{v-u' + \frac{i}{2}}{v-u'-\frac{i}{2}} \, , \qquad S^{31}(w,u') \eqsp \, \frac{w-u' + \frac{i}{2}}{w-u'-\frac{i}{2}} \, .
\eeq
Due to the underlying $so(6)$ symmetry, two level-2 or two level-3 magnons, respectively, scatter over each other with phases $S^{22}, S^{33}$, functionally identical to \eqref{S11}.

A concise and readable derivation of higher-level creation amplitudes and phases in a similar example is presented in \cite{beisertSu22}. This discussion is in fact about one wing algebra with three step operators. The second and third level are reached by supersymmetries differing in the Lorentz index. When both supersymmetries act on the same level-1 magnon, the top excitation is created. We can equally view this as $Q L Q$ with a Lorentz generator $L$ between two equal supersymmetry generators $Q$, all acting on the same level-1 site. Hence we have a \emph{composite}, or regarding the $Q$ generator, \emph{double} excitation.

\paragraph{Double excitations in the nested Bethe ansatz.}

The present note is concerned with double excitations, although in all our examples the excitation numbers decrease towards higher levels. The $so(6)$ sector displays the phenomenon too: the quadruple action $R_3^2 R_2^1 R_4^3 R_3^2$ maps $Z \rar \bar Z \eqsp \phi^{12}$. In other words, $\{n_1,n_2,n_3\} \eqsp \{2,1,1\}$ excitations at any one site yield $\bar Z$, where $n_i$ counts the number of level-$i$ excitations. 

In a state with $\{2,1,1\}$ excitations, generically the two secondary magnons sit on top of one level-1 excitation each, yielding $|\ldots Y \ldots \bar Y \ldots\ra$. This comes with the two creation amplitudes, say, $f^{21}(v,u_1) f^{31}(w,u_2)$, a product which is (symbolically) of degree $O(u^{-2})$. All $S$-matrices and also the momentum factors $e^{i p}$ are of overall degree 0 in rapidities and thus they do not change this estimate. By contrast, in the quadruple composite excitation $\bar Z$ the ansatz yields the longer product $f^{21}(v,u_1) f^{21}(v,u_2) f^{31}(w,u_1) f^{31}(w,u_2)$ of order $O(u^{-4})$. Comparing to dilatation-operator eigenstates obtained from direct diagonalisation, we observe that ket states with $\bar Z$ pick up the wrong coefficient using our nested ansatz as it stands. By homogeneity, we infer that a second-order numerator polynomial $p(\{u_1,u_2\},v,w)$ is missing. Matching on known eigenstates allows us to fix $p$ by polynomial interpolation and the result is
\beq
p(\{u_1,u_2\},v,w) \eqsp (u_1-u_2)(u_1-u_2-i) \, . \label{pPol}
\eeq
Notice that this expression is not symmetric under exchange of $u_1$ and $u_2$. Thus, level-1 rapidities are ordered for double excitations and, e.g., if the level-1 excitations are initially put onto the chain with rapidities labelled in ascending order $\{u_1, u_2, u_3 \ldots\}$, then we have to consider $u_i$ left of $u_j$ for $i \, < \, j$ even if they occupy the same lattice site when forming a quadruple excitation. Surprisingly, we can build the Bethe ansatz distributing all level-1 magnons first, and afterwards level-2,3, although the sequence of lowering operations is $R_3^2 R_2^1 R_4^3 R_3^2$ or $R_3^2 R_4^3 R_2^1 R_3^2$ to arrive at $\bar Z$ from the vacuum.

Next, the nested Bethe ansatz is supposed to diagonalise a \emph{matrix picture} in which excitations of the four flavours $X, Y, \, \bar Y, \bar X$ scatter. As there are several flavours, the scattering involves a true matrix of phases, which is the internal $su(2)_L \otimes su(2)_R$ part of the full magnon symmetry $psu(2,2)_L \otimes psu(2,2)_R$ of \cite{beiStau}, to which the full symmetry algebra $psu(2,2|4)$ of the \nFour theory is broken by the choice of vacuum. In this picture, the field $\bar Z$ is excluded, and also one half of the fermionic excitations, namely $\psi^{1\alpha}, \psi^{2\alpha},\bar \psi_3^{\dot \alpha}, \, \bar \psi_4^{\dot \alpha}$ and the chiral and antichiral field strengths $F^{\alpha\beta}, \tilde F^{\dot \alpha \dot \beta}$ are lost.

In \cite{higherRank} a possibility to mend this appeared for a $psu(1,1|2)$ sector with e.g.\ left and right fermions $\psi^{42}, \bar \psi_1^{\dot 2}$ as wing excitations on top of $X$, much like the scalars $Y, \bar Y$ in the $so(6)$ case. When the fermions occupy the same lattice site, they form an excitation $(D X)$ with an extra creation amplitude \eqref{pPol} in the nested Bethe ansatz. Yet, in that paper we were interested in computing tree-level three-point functions by the hexagon formalism \cite{BKV} which is most naturally formulated in the matrix picture. We therefore matched the latter with the Bethe wave function of the nested $psu(1,1|2)$ ansatz, starting with four wave functions with the initial flavour ordering
\beq
\{\psi^4, \bar \psi_1\} \, , \ \{\bar \psi_1, \psi^4\} \, , \ \{D, X\} \, , \ \{X, D\} \, .
\eeq
Importantly, in all four wave functions the first magnon has rapidity $u_1$ and the second $u_2$. For these we take over the level-1 rapidities of the relevant solution to the nested Bethe equations. Every wave function is given an a priori unknown amplitude $g_{\psi^4 \bar \psi_1}$ etc. We let the wave functions freely evolve like the Bethe state in an $su(2)$ sector, although the scattering is more complicated when magnon 1 overtakes magnon 2: $\psi^4 \bar \psi_1$ can scatter by the usual \emph{transmission} and \emph{reflection} terms (the ordering of flavour is flipped or not), but they can also become $D X$ or $X D$, mostly as two magnons on different sites. It is possible to analytically determine $g_{\psi^4 \bar \psi_1} \ldots $ comparing the ket states of the two ans\"atze without double excitations. Furthermore, we can match exactly introducing extra matrix-picture creation amplitudes
\beq
-e_{\psi \bar \psi} \eqsp e_{\bar \psi \psi} \eqsp e_{D X} \eqsp e_{X D} \eqsp U_{12} \, , \qquad U_{12} \eqsp \frac{u_1-u_2}{u_1-u_2-i} \label{extraF}
\eeq
for $|\ldots (DX) \ldots \ra$ arising from each of the four cases.

Remarkably, one can now compute structure constants modifying the hexagon formalism only minimally \cite{higherRank}: one builds \emph{entangled states} \cite{tailoring,BKV} (cf.\ \eqref{psi12XBX} below) from each wave function simply dragging along the relevant $g$ coefficient as a normalisation. The computations make no further reference to the local structure of the states --- the double excitations are swiped under the carpet again. Yet, thanks to their existence in the matrix picture, a new possibility appears: also in the hexagon approach it should be possible to put more than $L$ excitations onto a chain of length $L$! It is in this way that one may hope to realise the Lagrange operator.
 
Let us circle back to the $so(6)$ ansatz. We can interpret the quadruple excitation as
\beq
\bar Z \eqsp X + \bar X \eqsp Y + \bar Y \label{makeBar}
\eeq
because both double excitations provide a total of $\{2,1,1\}$ magnons\footnote{The oscillator picture of \cite{oneLoopDila} vindicates this too.}. On the other hand, away from coincidence at the same site, $X, Y, \, \bar Y, \bar X$ are exactly the excitations we can use in the matrix picture of \cite{beiStau}. In this note we ask the question whether we can always match the Bethe wave of the nested ansatz for a sector of the AdS/CFT integrable system onto a matrix picture with double excitations, if at the expense of introducing extra creation amplitudes. The answer of our study is affirmative! While we do not provide a general proof of the equivalence, our set of examples involves all possible double excitations\footnote{Here we do not refer to the effect of (possibly several) derivatives at the same site or on top of other magnons. With derivatives only, an effect shows at subleading orders in the coupling \cite{ES}.} showing  that \emph{all} extra creation amplitudes must have the form \eqref{extraF}.

\paragraph{Example: $L=2$ Konishi singlet.}

To illustrate the procedure let us consider the Konishi singlet at length $L=2$. Solving the $so(6)$ Bethe equations for $L \eqsp 2$ and excitations numbers $\{2,1,1\}$, one obtains a primary state with rapidities $-u_1 \eqsp 1/\sqrt{12} \eqsp u_2, \, v \eqsp 0 \eqsp w$. We stress that the equivalence of the two descriptions should take place off shell and without imposing cyclicity on the ket states. Exactly as in the $psu(1,1|2)$ example in \cite{higherRank}  the coefficients $g_{X \bar X}, \, g_{\bar X X}, \, g_{Y \bar Y}, \, g_{\bar Y Y}$ for the four possible wave functions can be fixed relying on the comparison of ket states without double excitations. In the case at hand, the result is
\begin{eqnarray}
g_{X \bar X} & = & f^{21}(v,u_2) \; f^{31}(w,u_2) \, S^{21}(v,u_1) \, S^{31}(w,u_1) \, , \nonumber \\
g_{\bar X X} & = & f^{21}(v,u_1) \; f^{31}(w,u_1) \, , \label{kon1ana} \\
g_{Y \bar Y} & = & f^{21}(v,u_1) \; f^{31}(w,u_2) \, S^{31}(w,u_1) \, , \nonumber \\
g_{\bar Y Y} & = & f^{21}(v, u_2) \; f^{31}(w, u_1) \ S^{21}(v,u_1) \, . \nonumber 
\end{eqnarray}
In the Konishi problem the remaining two constraints from $| Z \bar Z \rangle$ and $| \bar Z Z \rangle$ are equivalent. For $so(6)$ both particles are bosons so that we do not expect a sign difference between such amplitudes for $X \bar X$ and $\bar X X$ (the same for $Y, \bar Y$). We can then solve by the very ansatz \eqref{extraF} finding the signs
\beq
e_{X \bar X} \eqsp e_{\bar X X} \eqsp e_{Y \bar Y} \eqsp e_{\bar Y Y} \eqsp - U_{12} \, . \label{ampBZ}
\eeq
Upon normalising the state by \cite{Korepin}
\beq
\sqrt{L \, {\cal G} \, \prod \left(u_i^2 + \frac{1}{4}\right) \prod_{i<j} A_{ij}}~, \label{normalNorm}
\eeq
with $\mathcal G$ being the Gaudin determinant and $A_{ij}=S^{11}(u_i,u_j)$,
both descriptions yield the unit-norm singlet operator $\tr(X \bar X + Y \bar Y + Z \bar Z)/\sqrt{3}$ on shell. Note that
the $g$'s become equal up to signs: 
\beq
g_{X \bar X} \eqsp g_{\bar X X} \eqsp -g_{Y \bar Y} \eqsp -g_{\bar Y Y} \eqsp \frac{(-1)^{1/3}}{4 \, \sqrt{3}}~. \label{kon1Coef}
\eeq
A more comprehensive discussion of the idea of importing $g$ coefficients from the nested Bethe ansatz into the matrix picture can be found in \cite{higherRank}.  

\paragraph{Higher-rank excitations in the hexagon formalism.}

Let us provide some detail on how to use these ideas in the computation of structure constants by the hexagon formalism \cite{BKV}. In this formalism, a three-point function is cut into two halves, one back and one  front hexagon. To this end, the three single-trace operators realised as Bethe states also have to be cut: Their three back halves end up on every other edge of the back hexagon, the three front halves on those of the front hexagon. In the spectrum problem, two excitations scatter by a product of two $psu(2|2)$-invariant scattering matrices \cite{beisertSu22} taking out a common normalisation, while in the hexagon there is scattering by only one such matrix with a different overall scaling. A contraction rule then reads out permitted terms.

The hexagon formalism is therefore akin to the matrix picture of the spectrum problem and can be extended quite naturally to the higher-rank case by realising the three single-trace operators as  multi-component wave functions. While their direct construction in the matrix picture can be rather complicated, one can obtain them more easily by fixing the normalisation coefficients $g$ of each component in this wave function by matching with the corresponding nested Bethe state \cite{higherRank} as described above. 
These coefficients will contain the information originally carried by the secondary roots.
The central observation of \cite{higherRank} is that it is enough to provide the elementary excitations  $D, X, Y, \bar Y, \bar X, \psi^3, \psi^4, \bar \psi_1, \bar \psi_2$ with level-1 rapidities and to let them scatter on the hexagon -- the model tacitly contains the composite excitations. 

Hexagon amplitudes need to be normalised by
\beq
\sqrt{ \, {\cal G} \prod_{i<j} A_{ij}}
\eeq
instead of \eqref{normalNorm} in order to reproduce tree-level field-theory results for unit-norm eigenoperators. Note that for connected correlators, the field-theory combinatorics overcounts that of the hexagon approach by $\sqrt{L_1 L_2 L_3}$ as has been observed in many examples \cite{BKV,cushions}.

Let us have a brief look at the class of structure constants corresponding to three-point functions
\beq
\la {\cal K}^{L_1} \, {\cal O}^{L_2} \, {\cal O}^{L_3} \ra \, ,
\eeq 
where ${\cal O}^{L_2}$ and ${\cal O}^{L_3}$ are two vacua of lengths $L_2$ and $L_3$, while ${\cal K}^{L_1}$ are length-$L_1$ states with excitation numbers $\{4,2,2\}$ generalising the $L_1 \eqsp 2$ discussion above. The latter are the primaries (in the usual understanding based on $Q$ supersymmetry) of the multiplets containing the $su(2)$ sector two-magnon BMN operators \cite{bmn}. Their one-loop anomalous dimensions $\gamma_1$ are 
\beq
p \, = \, \frac{\pi \, n}{L_1+1} \, , \qquad \gamma_1 \, = \, 8 \sin^2 (p) \eqsp 6,4,5 \pm \sqrt{5},\ldots
\eeq
The corresponding generalisation of \eqref{kon1Coef} in closed form is
\beq
g_{X \bar X} \eqsp g_{\bar X X} \eqsp -g_{Y \bar Y} \eqsp -g_{\bar Y Y} \eqsp \frac{(-1)^{n/(L_1+1)}}{2 \, L_1 \, \sqrt{L_1+1}} \, .
\eeq

The remaining necessary input for the hexagon computation is the \emph{entangled state} \cite{tailoring,BKV} for the non-trivial operators ${\cal K}^{L_1}$. We have four wave functions with initial magnon positions $\{X_1 \bar X_2\}, \{\bar X_1 X_2\}, \{Y_1 \bar Y_2\}, \{\bar Y_1 Y_2\}$, where the subscript indicates the rapidities $u_j \, = (-1)^j/2 \, \cot(p/2)$. Each of the wave functions is cut into two parts, the first of which has length $l_{12} \eqsp (L_1+L_2-L_3)/2$ in the range $0 \ldots L_1$. By way of example, we  state the explicit split form in the first case:
\begin{eqnarray}
\Psi(L_1, \{X_1,\bar X_2\}) & \equiv & \Psi(l,\{X_1,\bar X_2\}) \, \Psi(\tilde l, \{\}) - e^{i \, p_2 \, l} \, \Psi(l,\{X_1\}) \, \Psi(\tilde l, \{\bar X_2\}) - \label{psi12XBX}  \\
&& e^{i \, p_1 \, l} \left[ \quad \frac{(AB_{12}^-)^2}{A_{12}} \ \Psi(l, \{\bar X_2\}) \, \Psi(\tilde l, \{X_1\}) +  \quad \frac{(AB_{12}^+)^2}{A_{12}} \ \Psi(l,\{X_2\}) \, \Psi(\tilde l, \{\bar X_1\}) \right] - \nonumber \\ 
& & e^{i \, p_1 \, l} \left[ \frac{AB_{12}^- AB_{12}^+}{A_{12}} \ \Psi(l, \{\bar Y_2\}) \ \Psi(\tilde l, \{Y_1 \, \}) +  \frac{AB_{12}^- AB_{12}^+}{A_{12}} \ \Psi(l,\{Y_2 \, \}) \ \Psi(\tilde l, \{\bar Y_1\}) \right] + \nonumber \\ 
&& e^{i (p_1+p_2) l} \, \Psi(l,\{\}) \, \Psi(\tilde l, \{X_1,\bar X_2\}) \, + \, O(g)\, , \nonumber 
\end{eqnarray}
where we have used the abbreviations
\beq
l \eqsp l_{12} \, , \qquad \tilde l \eqsp L_1 - l_{12} \, ,  \qquad AB_{12}^{\mp} \eqsp \frac{A_{12} \mp B_{12}}{2} 
\eeq
with $A_{12}, \, B_{12}$ the $psu(2|2)$ $S$-matrix elements according to \cite{beisertSu22}. One can explicitly check --- including ket states with the double excitation $\bar Z$ --- that \eqref{psi12XBX} is an identity. At this point our toolbox for the tree-level hexagon computation is complete.

Notice that the $C_{12}$ element of the $S$ matrix \cite{beisertSu22} can in principle scatter two bosons into two fermions where that is allowed by the quantum numbers characterising the state. This would introduce coupling-constant dependent terms into the above, the omission of which we have indicated by the symbol $O(g)$. However, in an $so(6)$ sector the scattering does not factorise beyond tree level so that we \emph{must} limit the analysis here presented to $O(g^0)$. In principle, it should be possible to recover the ${\cal K}^{L_1}$ states from the higher-loop nested Bethe ansatz of \cite{beiStau}, albeit for much higher excitation numbers. If we managed to extract the coefficients of the matrix ansatz and the level-1 rapidities from there, our approach should be capable of reproducing loop effects, both in the spectrum problem and for structure constants. Further sources of $g$ dependence would be the replacement of the rapidities by Zhukowsky variables $x^\pm$ \cite{beiStau}, the BES phase \cite{BES}, the loop corrections introduced into the rapidities upon solving the nested Bethe equations, and finally gluing corrections \cite{BKV}. 

Returning to the tree problem, due to the various considerations regarding normalisations, we expect the relation
\beq
{\cal A}_\text{QFT} \eqsp \left(u^2 + \frac{1}{4}\right) L_1 \sqrt{L_2 L_3} \ {\cal A}_\text{hexagon}\, , \label{agree}
\eeq
where $u$ denotes one of the two level-1 rapidities. Apart from the latter, the hexagon amplitudes only depend on the quantum number $l_{12}$. The amplitude for $l_{12} \eqsp 0$ manifestly vanishes and
\beq
{\cal A}_\text{hexagon}^{l_{12}=1}(-u,u) \eqsp \frac{8 \, g_{X \bar X} \, u}{(u - \frac{i}{2})(u+\frac{i}{2})^2} \eqsp \frac{\sqrt{3}}{2}, \, \frac{\sqrt{2}}{3}, \, \frac{1}{4}, \, \ldots \label{hexaOut}
\eeq
for the $L_1 \eqsp 2, 3$ and the two $L_1 \eqsp 4$ operators. A field-theory check is very simple in this case: Let us put the unit-norm ${\cal K}^{L_1}$ operator into a three-point function with unit-norm vacua $\tr(Z^{L_2})/\sqrt{L_2}$ and $\tr(\bar Z^{L_3})/\sqrt{L_3}$. On the back of an envelope we find, putting $x_1, x_2, x_3$ to $0,1,\infty$ and focusing on the leading-$N$ coefficient,
\beq
{\cal A}_\text{QFT} \, = \, c_{\bar Z}(L_1) \ \sqrt{L_2 L_3}  \label{qftOut}
\eeq
if the correlator exists. Here $c_{\bar Z}(L_1)$ is the coefficient of the $\tr(\bar Z Z^{L_1-1})$ term in the operator. The numbers \eqref{qftOut} immediately satisfy \eqref{agree} employing the values listed in \eqref{hexaOut}.

Beyond $l_{12} \eqsp 0,1$, for every Bethe solution the hexagon computations yield an $l_{12} \rar L_1 - \l_{12}$ symmetric set of mostly non-vanishing numbers whose field-theory interpretation is somewhat obscure. They can in fact be reproduced from Wick contractions for a good range of examples using the shifted vacua \cite{BKV}
\beq
\cO^L \eqsp \frac{\tr(\hat Z^L)}{\sqrt{L}} \, , \qquad \hat Z(a) \, = \, Z + a \, (Y - \bar Y) + a^2 \bar Z
\eeq
at points 2,3, where the operator is put at the position $a$ along the $x_2$ axis. A stumbling block is the quickly growing combinatorics because $\cO^L$ now has $4^L$ terms as opposed to solely one in the direct test.

\paragraph{Lagrangian insertion method.}

In field-theory computations in the \nFour model \cite{nilpotent} we have much profited from the method of \emph{Lagrangian} insertions: differentiation in the coupling constant inserts an additional, integrated-over Lagrange operator into an $n$-point function. At least for higher-point functions of half-BPS operators, the trick gives a very convenient way of constructing higher-loop integrands without drawing upon Feynman graphs \cite{hidden}. In the context of the hexagon formalism, the hope is that this procedure can compute non-planar corrections to the spectrum by inserting the Lagrange operator into a two-point function on a tessellation of a torus (or possibly higher-genus) diagram \cite{colourDressed,handles1,doubleTorus}, and potentially also to sidestep, or at least to simplify, the evaluation of gluing contributions \cite{shotaThiago1,fivePoints}.

The Lagrangian of \nFour super Yang-Mills theory has the full supersymmetry only on shell, so when the equations of motion are satisfied. In the next section we state a form of the on-shell Lagrangian\footnote{It can be rewritten in many ways by using the equations of motion or by adding topological terms.} which is the chiral Yang-Mills Lagrangian with two higher-order admixtures: a Yukawa term at linear order in the coupling constant, and the quartic scalar superpotential at $O(g^2)$. In the \emph{integrability} philosophy \cite{beiStau} admixtures are usually not mentioned; their effect is captured by the coupling-constant dependence of the Zhukowsky variables, and loop corrections to the rapidities of a given leading-order solution of the Bethe equations\footnote{We are grateful to A. Sfondrini for a discussion on this issue.}. In the following we study the planar mixing problem of these operators at leading order in the coupling constant. Doing so we will learn some facts about descendents, in particular in the matrix picture.

This article is organised as follows: in the next section we briefly comment on the operator mixing between the $SU(4)$ singlet operators of naive scaling dimension 4, in analogy to the situation in the \textbf{10} of $SU(4)$ at weight 3. Section \ref{threeParts} is the main part of the article: we discuss in turn how the these operators can be realised by a multi-component wave function comparing to the nested Bethe ansatz.  In passing we have to fix quite a number of creation amplitudes for double excitations. In the nested ansätze these can become quite fanciful, in the matrix picture there is complete universality. In the example of the Yukawa term we learn how we could later come to grips with the Yang-Mills action. Finally, we present some conclusions.

\section{The on-shell Lagrangian} \label{secL}

Appendix A.2 of \cite{superCorAmp1} presents a form of the on-shell Lagrangian of \nFour super Yang-Mills which has been tried and tested in Feynman diagram computations with Lagrangian insertions:
\beq
\cL \, = \, \tr\left( -\frac{1}{2} F_{\alpha \beta} F^{\alpha \beta} + \sqrt{2} g \, \psi^{\alpha I} [\phi_{IJ}, \psi^J_\alpha] - \frac{1}{8} g^2 \, [\phi^{IJ},\phi^{KL}][\phi_{IJ},\phi_{KL}] \right)~. \label{ourL}
\eeq
In contracting $\cL$ on e.g.\ scalar operators, vertices may be required at $O(g^2)$ but the diagrams will then be so that the integrations break due to antisymmetrised derivatives \cite{antisym}. In the integrability picture that ought to be captured by the solution of the mixing problem corresponding to the leading-order eigenstate $\tr(F^2)$. Yet, in order to gain experience with highly-excited states on the hexagon, we will study all $SU(4)$-singlets of classical dimension $4$ in the following.
The material of \cite{higherRank} gives us a handle to do so. 

The specific linear combination \eqref{ourL} is finite. To better understand this, let us draw a parallel to the weight-3 mixing problem in the \textbf{10} of $SU(4)$, see \cite{koniAno}. In the highest-weight component there are the two operators
\beq
\tr(\psi^4 \psi^4) \, , \qquad \tr(X \, [Y, \, Z])
\eeq
that may mix. Taking a double derivative $(Q_3)^2$ of the length 2 vacuum $\tr(Z^2)$, we produce the double fermion operator by acting on either site of the vacuum with one $Q_3$, while the second operator comes from acting with both supersymmetry generators on the same site and employing the chiral fermion equation of motion. The result is an operator of the form
\beq
\tr(\psi^4 \psi^4)  - 4 \, g \, \tr(X \, [Y, \, Z]) \, . \label{O10}
\eeq
As a supersymmetry descendent of the vacuum this operator should be protected. Then it must be so that the two-fermion term has vanishing one-loop anomalous dimension. At higher orders of a two-point function, divergences and the associated logarithms cancel between the contributions from the two operators. Indeed, divergences connected to the anomalous dimension of the operator will not occur in any correlation function.

On the other hand, the $\tr(X \, [Y,Z])$ state is the leading part of a descendent of the Konishi singlet:
\beq
(\bar Q^4)^2 \, \tr(\phi^{IJ} \phi_{IJ}) \sim g \, \tr(X \, [Y, \, Z])~.\label{K10}
\eeq
Once again, the (here antichiral) fermion equation of motion is at the root of this. The operator inherits the one-loop anomalous dimension 6 of its primary. It receives an $O(g^2)$ admixture $g^2 N/(32 \pi^2) \, \tr(\psi^4 \psi^4) + \ldots$ which one can see imposing orthogonality. Crucially, this does \emph{not} follow from classical supersymmetry acting on the Konishi singlet, or at least one needs a point splitting regulator as in Konishi's original argument or to appeal to the $U(1)$ anomaly \cite{koniOr}. Therefore the term is called the Konishi \emph{anomaly}.

The two complete eigenstates \eqref{O10} and \eqref{K10} should both be visible as solutions of an appropriate system of (nested) Bethe equations. The two-fermion operator is most easily realised in an $su(1|1)$ sector. Since it is a descendent of the length-2 vacuum, we expect two infinite level-1 rapidities. On the other hand, as the fermion equation of motion is not manifest in the  Bethe ans\"atze, the $\tr(X \, [Y, \, Z]) + \ldots$ operator appears as a primary state at length 3. Indeed, in the $so(6)$ ansatz sketched in Section 1 we find the usual Konishi level-1 rapidities $\pm 1/\sqrt{12}$ and a secondary one $v \eqsp 0$.

Likewise, in $\cL$ the chiral Yang-Mills action on its own will be one-loop protected but it needs the specific admixtures in \eqref{ourL} to remain finite at higher orders in the coupling. The operator \eqref{ourL} --- or anything related to it by equations of motion or surface terms --- arises as the quadruple susy trafo $Q_3^2 \, Q_4^2 \, \tr(Z^2)$. For the leading term $\tr(F^2)$ we thus expect four infinite level-1 rapidities, and as well as several secondary ones since we need to change flavour on two of the supersymmetry variations and to raise two Lorentz indices. To extract this from the gradings proposed in \cite{beiStau} would introduce many secondary roots. Rather using $Q_3^2$ ($\alpha \eqsp 2$) as a primary excitation we can employ $L_2^1, \, R_4^3$ on a left and right wing, respectively. It follows that this is an $L \eqsp 2$ state with excitation numbers $\{4,2,2\}$. We should assume \emph{all} associated Bethe roots infinite.

Because the chiral field strength tensor $F^{\alpha \beta}$ is symmetric in its indices and the trace is cyclic, antisymmetrisation $\tr(F^{\alpha [\beta} F^{\gamma ] \delta})$ does imply antisymmetry on $\alpha, \delta$. Therefore we only find two dilatation operator \cite{oneLoopDila} eigenstates in this sector: the doubly symmetrised one has one-loop anomalous dimension 6 and the doubly contracted one of interest to us is one-loop protected as anticipated.

Second, scanning for one-loop eigenstates by means of the dilatation operator, the superpotential term 
\beq
\tr\left( [\phi^{IJ},\phi^{KL}][\phi_{IJ},\phi_{KL}] \right)
\eeq
amusingly turns out to be a sum of two primaries with anomalous dimensions $\frac{1}{2}(13 \pm \sqrt{41})$ at $g^0$. Appealing to cyclic symmetry, the operator has 24 ket states differing by flavour. Their coefficients in the two eigenstates are
\beq
| A A \bar A \bar A \ra \ : \ 1 \mp  \frac{11}{\sqrt{41}} \, , \qquad | A B \bar A \bar B \ra : 2 \mp \frac{2}{\sqrt{41}} \, , \qquad \text{else} \ : \ -1 \mp \frac{9}{\sqrt{41}} \label{cofSPP}
\eeq
for two scalars $A \, \neq \, B, \, A \, \neq \, \bar B$.

Third, let us have a look at the fermion term $\tr(\psi^{\alpha I} [\phi_{IJ}, \psi^J_\alpha])$. First, dropping the commutator, field theory allows us to write 12 distinct terms pinching the $SU(4)$ indices (not the Lorentz contraction): there are six distinct scalars and two cyclic orderings of the terms. The eigenspectrum of the dilatation operator is again a little surprise: separately for every scalar, the commutator terms have anomalous dimension 6. The other combinations have $\gamma_1 \, \in \, \{4,0\}$. Four of the six $\gamma_1 \eqsp 6$ operators will arise from $(Q_J)^2 (\bar Q^J)^2$ (so sum over $J$) acting on the Konishi singlet, for the other two we lack an interpretation at this point. Note that $(\bar Q)^2$ is applied first to generate the antichiral equation of motion. $Q^2$ then yields a chiral Yukawa-like term. In the opposite order we would find antichiral fermion terms.

\section{Operators in the Lagrangian mixing problem from the Bethe ansatz} \label{threeParts}


In the following we study the operators appearing in the on-shell Lagrangian mixing problem, and find the leading-order conformal eigenstates from the Bethe ansatz. In particular, we expand on the previous discussion of the mixing of operators schematically of the type $\tr(F^2),\tr(\psi\phi\psi)$ and $\tr(\phi^4)$, and realise them as conformal eigenstates in the matrix picture by matching onto the corresponding nested Bethe state. At tree-level these three sectors decouple, and there is no restriction on the choice of grading for the nested Bethe ansatz for the $psu(2,2)_L \otimes psu(2|2)_R$ symmetry of the problem \cite{superSpinChain}. To the end of minimising the number of Bethe roots we adapt the grading to each of the three cases:
\begin{center}
\quad \phantom{w \,} \includegraphics[height = 2.75 cm,page=1]{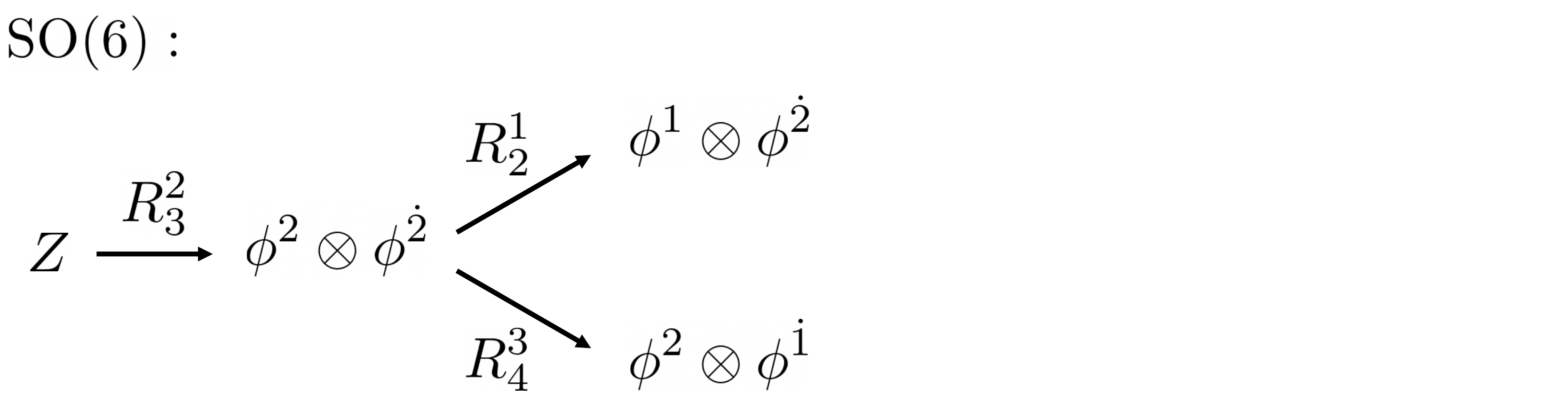}\\
\includegraphics[height = 2.75 cm,page=2]{figures.pdf}\\
\phantom{w \,} \includegraphics[height = 2.75 cm,page=3]{figures.pdf} \\[3 mm]
Figure 1: Bethe ans\"atze for the three operators in $\cL$
\end{center}
The breaking of the full $psu(2,2|4)$ symmetry into the product of a left- and a right-wing subalgebra is a result of the choice of vacuum, for us $Z \, = \, \phi^{34}$. In the usual formalism this impedes the use of $\bar Z$ and one half of the fermions, as well as the chiral and anti-chiral field strength tensors as excitations on the chain. We have discussed in Section 1 how to recover $\bar Z$ as a double excitation; in the following we will see that the same is possible for $\psi^1, \psi^2, F$ or similarly on the other wing $\bar \psi_3, \bar \psi_4, \tilde F$.

\subsection{Superpotential-type terms} \label{potentiallySuper}

Let us study any one term in the superpotential, for simplicity we begin with an example not involving $\bar Z$. Specifically, we study length-4 states with four excitations $X, X ,\bar X, \bar X$, i.e.\ excitation numbers $\{n_1,n_2,n_3\} \eqsp \{4,2,2\}$. The nested Bethe equations for $so(6)$ have solutions with the aforementioned energies (viz one-loop anomalous dimensions) $\gamma_1 \, = \, \frac{1}{2}(13 \pm \sqrt{41})$. For the top state (plus sign) the momentum carrying rapidities are 
\beq
u \, \in \, \left\{ \pm \frac{1}{2} \sqrt{\frac{1}{15} \left(3+2 \sqrt{41}+\sqrt{42 \sqrt{41}-82}\right)}, \, \pm \frac{1}{2} \sqrt{\frac{1}{15} \left(3+2 \sqrt{41}-\sqrt{42 \sqrt{41}-82}\right)}\right\} \label{topUs}
\eeq
and the secondary pairs are both ($v_i \eqsp w_j$ will not send the wave function to zero)
\beq
v,w \eqsp \pm \sqrt{\frac{2}{1 + \sqrt{41}}} \, . \label{topVWs}
\eeq
Here all rapidities are real. Flipping the sign of $\sqrt{41}$ makes all rapidities complex, but the Bethe equations are solved, and the energy is indeed $(13 -\sqrt{41})/2$. Both states are recognised as primaries by the finiteness of all roots.

Formulae \eqref{pPol}, \eqref{makeBar}, \eqref{ampBZ} are enough to match (i) the Bethe state of the nested $so(6)$ ansatz\footnote{Our standard normalisation \eqref{normalNorm} is not appropriate for these states. Curiously, for the real Bethe solution corresponding to the top operator even the phase comes out wrong. We have normalised  \eqref{gOutPlus}, \eqref{gOutMinus} such as to reproduce \eqref{cofSPP}.} onto the two dilatation-operator eigenstates \eqref{cofSPP}, and (ii) the matrix ansatz onto the nested one. The latter has 36 wave functions with the initial orderings
\beq
\{X, Y, \, \bar Y, \bar X\} + 23 \; \text{permutations} \, , \qquad \{X, X, \bar X, \bar X\} + 5 \; \text{others} \, , \qquad \{Y, \, Y,  \, \bar Y, \bar Y\} + 5 \; \text{others} \, .
\eeq
For the top state these come with the coefficients
\begin{eqnarray}
g_{A A \bar A \bar A} & = & -1.5116253171132683576 + 1.7991398887412476048 \, i  \, , \nonumber \\
g_{A \bar A A \bar A} & = & -11.324250791414210813 - 9.4734590640941648254 \, i \, , \nonumber \\
g_{A \bar A \bar A A} & = & -1.1007652906361257446 - 8.7060249412316226459 \, i \, ,\label{gOutPlus} \\
g_{A B \bar A \bar B} & = & -4.3559300918324083552 - 1.2832870058018948922 \, i \, , \nonumber \\
g_{A B \bar B \bar A} & = & \phm 5.8675554089456767127 - 0.5158528829393527127 \, i \, , \nonumber \\
g_{A \bar A B \bar B} & = & \phm 5.4566953824685340998 + 9.9893119470335175381 \, i \, . \nonumber
\end{eqnarray}
for $A, B \, \in \, \{X, Y, \, \bar Y, \bar X\}, \,  A \, \neq \, B , \, A \, \neq \, \bar B$. 

The equality of nested and matrix Bethe state has to happen off shell and prior to imposing cyclicity on the ket states. There are 36 ket states without double excitations, so precisely as many as unknown coefficients. The system of equations has maximal rank in this way, whereas it is rendered underdetermined by the cyclic identification of equivalent kets.
Substituting the Bethe solution \eqref{topUs}, \eqref{topVWs} we can thus in principle exactly pin down all the coefficients. However, due to the proliferation of Bethe roots we were not able to analytically solve for the $g$'s as in the Konishi singlet example, cf.\ equations \eqref{kon1ana}.  Disappointingly, we could not extract exact results as in \eqref{topUs}, \eqref{topVWs} either due to the complicated square root structure of the rapidities. Numerically, the coefficients can, of course, be determined to any desired precision\footnote{We have worked with 200 digits of precision in {\tt Mathematica}.}. A quick test with PSLQ suggests that the numbers in \eqref{gOutPlus} cannot be decomposed over $\{1, \sqrt{41}\}$ as one might guess regarding \eqref{cofSPP}. Last, we remark that the creation amplitudes \eqref{pPol} and \eqref{ampBZ}, respectively, yield consistent results for the remaining 54 ket states with $\bar Z$.
For the bottom state we find in similar same way
\begin{eqnarray}
g_{A A \bar A \bar A} & = &  -0.24874753055421299045 + 1.79596710929995399815 \, i  \, , \nonumber \\
g_{A \bar A A \bar A} & = &  -0.06217033900405760969 + 0.44887233162292967920 \, i \, , \nonumber \\
g_{A \bar A \bar A A} & = &  -0.06220342847390091424 + 0.44911123891728690209 \, i \, ,\label{gOutMinus} \\
g_{A B \bar A \bar B} & = &  \phm 0.12439031001202814750 - 0.89810300829715561052 \, i \, , \nonumber \\
g_{A B \bar B \bar A} & = &  \phm 0.12435722054218484296 - 0.89786410100279838763 \, i \, , \nonumber \\
g_{A \bar A B \bar B} & = &  -0.06218688153812723326 + 0.44899176937986870843 \, i \, . \nonumber
\end{eqnarray}

\subsection{Yukawa-type terms}

We now move on to Yukawa-type terms and will disregard the standard gradings \cite{beiStau} for this exercise as well, because we know how $\tr(X \, [Y,\, Z])$ can be realised as a primary state with three roots (two momentum carrying and one secondary), see Section \ref{secL}. Furthermore, since $\phi^{34}$ was singled out as a vacuum, only the operations $Q_J^2 (\bar Q^J)^2 {\cal K}_1, \, J \eqsp 3,4$ are appropriate. The $J \eqsp 3$ version yields the excitations $\bar Y, \bar X$ and thus a state with excitation numbers $\{2,1,2\}$. This is a double descendent of the $J \eqsp 4$ operator (equivalently, note it is obtained by two $R_4^3$ step operators). Acting further with the lowering operators, we can thus only hope to find one out of the six $\gamma_1 \eqsp 6$ eigenstates. It will in fact be the singlet as we desire, i.e.\ the sum of all six terms from above! All primary and secondary rapidities brought in by the lowering should be infinite.

In the second panel of Figure 1 we sketch the necessary part of a suitable nested Bethe ansatz: if we label the $so(6)$ nodes at the root of the tree by level 1,2,3, the fourth node with raising operator $Q_1^2$ ($\alpha \eqsp 2$ is implied) is connected to node 2, the fifth with generator $L_2^1$ to the fourth. The creation amplitudes and scattering matrices of the $so(6)$ part are as before. At level $4, 5$ we call the auxilliary rapidities $x_j, y_k$, respectively. The additional creation amplitudes and scattering phases are
\begin{eqnarray}
&& f^{42}(x,v) \eqsp \frac{1}{x-v-\frac{i}{2}} \, , \qquad S^{42}(x,v) \eqsp \frac{x-v+\frac{i}{2}}{x-v-\frac{i}{2}} \, , \qquad S^{44}(v_1,v_2) \, = \, -1 \,  \label{level45bits} \\
&& f^{54}(y,x) \eqsp \frac{1}{y-x+\frac{i}{2}} \, , \qquad S^{54}(y,x) \eqsp \frac{y-x-\frac{i}{2}}{y-x+\frac{i}{2}} \, , \qquad S^{55}(y_1,y_2) \eqsp \frac{y_1 - y_2 + i}{y_1 - y_2 - i} \, . \nonumber
\end{eqnarray}
To understand the excitation numbers needed for the Yukawa term, we only need to inspect any one term in the mixing problem, for simplicity consider $\psi^3 \psi^4 \bar Z$. As usual, to obtain $\bar Z$ we need $R_3^2 R_4^3 R_2^1 R_3^2$, so $\{n_1, n_2, n_3, n_4, n_5\} \eqsp \{2,1,1,0,0\}$. Second, the fermion on the left/upper branch of the Dynkin diagram is $\psi_4$. To reach $\psi^{42}$ we require $Q_1^2 R_2^1 R_3^2$. Hence momentarily putting supersymmetry or Lorentz transformations aside, another level-1 and level-2 magnon are needed. Third, $\psi^3$ requires a right magnon as well: here the sequence of raisings is $R_4^3 Q_1^2 R_2^1 R_3^2$. Finally,  $n_5 \, = \, 1$ because there is one Lorentz contraction. In total, our vector of excitation numbers is
\beq
\{n_1,n_2,n_3,n_4,n_5\} \eqsp \{4,3,2,2,1\} \, 
\eeq
in the aforementioned ordering of raising operators, i.e.\ the nodes 1 \ldots 5 correspond to $\{ R_3^2, R_2^1, R_4^3, Q_1^2, L_2^1 \}$. 

Before seeking Bethe solutions, let us comment on multiple excitations: the \emph{lowest} fermion in terms of the $R$ index we have met so far was $\psi^3$. Yet, we also need the term $\tr(\psi^1 \psi^2 Z)$, say. Note that  $\psi^2 \eqsp R_3^2 \, \psi^3, \, \psi^1 \eqsp R_2^1 \, \psi^2$. In both cases, there is a double level-1 rapidity, but for $\psi^1$ also $R_2^1$ must be doubled. In analogy to \eqref{makeBar}, the rules
\beq
\psi^1 \eqsp \psi^3 + Y \eqsp \psi^4 + \bar X \, , \qquad \psi^2 \eqsp \psi^3 + X \eqsp \psi^4 + \bar Y \label{rules2}
\eeq
should be valid. Further, counting excitation numbers (or by the oscillator picture from \cite{oneLoopDila}) one can also check
\beq
F \eqsp \psi^3 + \psi^4 \, . \label{rules3}
\eeq
Letting the twelve excitations freely spread over the three sites we do indeed generate all possible fermion terms; each of these has the same total excitation numbers. The rules \eqref{makeBar}, \eqref{rules2}, \eqref{rules3} will guarantee this in the matrix picture.

Explicitly, the excitation numbers of the different fermions are
\beq
\psi^4 \, : \, \{1,1,0,1,0/1\} \, , \quad \psi^3 \, : \, \{1,1,1,1,0/1\} \, , \quad \psi^2 \, : \, \{2,1,1,1,0/1\} \, , \quad \psi^1 \, : \, \{2,2,1,1,0/1 \} \, , \label{fermVecs}
\eeq
while the field strength $F$ must have excitation numbers $\{2,2,1,2,0/1/2\}$ being a sum $\psi^3 + \psi^4$. Apart from those 5-tuples characterising the scalars, other combinations of $\{n_i\}$ than in the above are put to zero.

The first non-standard case in the list \eqref{fermVecs}  is $\psi^2$ with a double level-1 excitation. Likely, we have the same situation as for $\bar X$ since the Lorentz index can be effectively removed putting $n_5 \eqsp 0$. Indeed, this is confirmed by matching on dilatation-operator eigenstates obtained by direct diagonalisation. On the other hand, for $\psi^1$ the nested Bethe ansatz runs up the eight creation amplitudes
\beq
f^2(v_1,u_1) f^2(v_1,u_2) f^2(v_2,u_1) f^2(v_2,u_2) f^3(w_1,u_1) f^3(w_1,u_2) f^4(x_1,v_1) f^4(x_1,v_2)
\eeq
instead of a total of four amplitudes in e.g.\ a ket state with $\psi^4$ and $\bar X$ at different sites. Hence, by the intuition leading to \eqref{pPol}, we seek an extra fourth-order numerator polynomial $q$ which we will again try to pin down by interpolation. To avoid complications with infinite roots, let us compare to a convenient set of primary states. These should have as few Bethe roots as possible, yet they must comprise the $\psi^1$ fermion. States with a single $\psi_1$ (and what is related by spreading out the magnons along the chain) are apparently always descendents, at least we have not found Bethe solutions without infinite roots.

The next best choice is then $\tr(Z^{L-2} X \psi^1) + \ldots$ with excitation numbers $\{3,2,1,1,0\}$. At lengths $L \eqsp 4 \ldots 7$ we find two types of solutions: type (i), say, has $v_2 \eqsp 0 \eqsp w_1$, while type (ii) has $v_2 \eqsp 0, \, v_1 \eqsp w_1 \eqsp 2 \, x_1$. As might be expected for three momentum-carrying rapidities, the root distributions for $\{u_1,u_2,u_3\}$ are all asymmetric (even though (i) type solutions with a vanishing $u$ exist) and therefore create parity pairs. The anomalous dimensions are
\vskip 0.2 cm
\begin{center}
\begin{tabular}{c|c|c} 
$L$ & type & $\gamma_1$\\
\hline
4 & (ii) & $15/2$ \\[1 mm]
5 & (i) & 8 \\[1 mm]
5 & (ii) & 6 \\[1 mm]
6 & (i) & 7 \\[1 mm]
6 & (ii) & $(25 \pm \sqrt{37})/4$ \\[1 mm]
7 & (i) & 6,\, 8 \\[1 mm]
7 & (ii) & 3.75979, \, 6.83525, \, 8.40496
\end{tabular}
\end{center}
The cases $5(ii)$ and $7(i)$ fall into eigenspaces of fairly large dimension and are therefore not useful for our purposes. In the other cases, we can clearly identify the corresponding pairs of eigenvectors of the dilatation operator \cite{oneLoopDila} by the value of $\gamma_1$ and re-arrange them into parity-even/odd combinations. To form the equivalent from the Bethe states, we enforce the absence of those kets without $\psi^1$ that are missing in the field-theory states\footnote{For the length-7 states this fails for the parity-odd parts which contain all possible kets. However, we may use orthogonality to the even parts (see \cite{doubleTorus}) to fix the odd linear combinations.}. Upon normalising the first existing ket state in the linear combinations to have coefficient 1, we can compare field theory and Bethe states. We observe immediate agreement in all terms barring for those with $\psi^1$. So the interpolation for $q$ may begin!

The creation amplitude $p$ in \eqref{pPol} displays a factor $(u_1-u_2)$. This reflects the fact that the entire wave function must vanish when there is a degeneracy between the rapidities of any given type. In $q$ we might assume the two factors $(u_1-u_2)(v_1-v_2)$ times a general second-order polynomial depending on all rapidities. Of course, for our choice of states, no information on the coefficients of terms with $v_2$ can be extracted. Some freedom remains even so. Higher-length solutions of the same type are unlikely to provide more information. Yet, excluding the rapidities $w_1,x_1$ from our intermediate result is possible and even leaves a one-parameter freedom. With hindsight we choose
\beq
q(\{u_1,u_2\},\{v_1,0\}) \, = \, -(u_1-u_2)(u_1-u_2-i) \, v_1 (v_1-i) \label{oriQ}
\eeq
whose obvious extension to general rapidity $v_2$ is
\beq
q(\{u_1,u_2\},\{v_1,v_2\}) \, = \, -(u_1-u_2)(u_1-u_2-i) (v_1 -v_2)(v_1-v_2-i) \, .
\eeq
In fact, this looks reasonable in the light of the $SO(6)$ symmetry in the $u,v,w$ sector; we could have built a Bethe ansatz with $R_2^1$ as a level-1 excitation, too, were it not for the choice of vacuum. 

Beyond the creation amplitudes in the nested ansatz, our $\tr (Z^{L-2} X \psi^1) + \ldots$ problem allows us to study
those for the double excitations \eqref{rules2} in the multi-component wave functions. The sign rules
\beq
e_{X \psi^3} \eqsp e_{\psi^3 X} \eqsp e_{Y \psi^3} \eqsp e_{\psi^3 Y} \eqsp e_{\bar X \psi^4} \eqsp e_{\psi^4 \bar X} \eqsp - e_{\bar Y \psi^4} \eqsp -e_{\psi^4 \bar Y} \eqsp U_{12} \label{fermCreate}
\eeq
allow for exact equality between nested and matrix ansatz. 

By construction, the field strength $F$ does not arise in the aforementioned set of primary states. The simplest suitable situation features the two fermions $\psi^{32},\psi^{42}$ (two equal Lorentz indices since $F$ is a symmetric tensor) and thus has excitation numbers $\{2,2,1,2,0\}$. We have not yet sought primary states of this type, possibly it will again be necessary to include other fields.

Returning to the Yukawa-type terms, let us comment on candidate Bethe solutions. As already mentioned, the Bethe vectors will vanish if any two rapidities at a given level are degenerate. Hence we cannot simply put rapidities to $\infty$ in some arbitrary way. A good idea when dealing with descendents is usually to introduce a twist regulator. Unlike in \cite{raduNiklas,higherRank}, this will not be defined on the level of the Lagrangian by a charge wedge, but rather we arbitrarily introduce twist $e^{i \, \beta \, m_j}$ at node $j$ of the Bethe ansatz; $m_0$ refers to such a parameter in the momentum equation. In conjunction with this, every rapidity is expanded in the order parameter $\beta$:
\beq
u_j \rar \frac{u_{j,-1}}{\beta} + u_{j,0} + u_{j,1} \, \beta + u_{j,2} \, \beta^2 + u_{j,3} \, \beta^3 + \ldots \
\eeq
and likewise for the auxilliary ones. The three known finite rapidities --- let us single out $u_3,u_4,v_3$ --- are expanded in the same manner assuming the $\beta^{-1}$ contribution to vanish. Barring for the one level-2 rapidity $v_3 \eqsp 0 + \ldots$ and the one level-5 case, there are even numbers of rapidities. Many solutions of the nested equations have pairs of rapidities of opposite sign (for  instance here $u_4 \eqsp 1/\sqrt{12} \eqsp - u_3$), a very useful simplification. Putting $x_1 \eqsp -x_2$ turns out to be quite restrictive. Relaxing this constraint we find, expanding the Bethe equations up to $O(\beta^1)$, 
\begin{eqnarray}
&& m_0 \eqsp m_1 \eqsp m_3 \eqsp 0 \, , \nonumber \\
&& u_{1,-1} \eqsp - u_{2,-1} \eqsp 1 \, , \quad u_3 \, = \,  -\frac{1}{\sqrt{12}} + O(\beta^2) \, = \, - u_4  \, , \nonumber \\
&& v_{1,-1} \eqsp - v_{2,-1} \eqsp - \sqrt{2} \, , \quad v_3 \, = \, 0 + O(\beta^2) \, , \label{all3} \\
&& w_{1,-1} \eqsp - w_{2,-1} \eqsp -\frac{1}{\sqrt{3}} \nonumber
\end{eqnarray}
universally, beyond which there are the three distinct solutions
\begin{eqnarray}
&& \{ m_{2} = \phm 0, \, m_{4} = -3, \, m_{5} = \phm 6, \ x_{1,-1} =  -\sqrt{\frac{10}{3}}, \, x_{2,-1} =  \sqrt{\frac{10}{3}}, \, y_{1,-1} = -\frac{5}{3} \} \, , \nonumber \\
&& \{ m_{2} = -3, \, m_{4} =  \phm 3, \, m_{5} =  \phm 3, \ x_{1, -1} = -1 - \frac{1}{\sqrt{3}}, \, x_{2, -1} = -1 + \frac{1}{\sqrt{3}}, \, y_{1, -1} = -\frac{4}{3} \} \, , \label{sixSolutions} \\
&& \{ m_{2} = -3, \, m_{4} =  \phm 6, \, m_{5} = -3, \ x_{1, -1} = -1 - \frac{1}{\sqrt{3}}, \, x_{2, -1} = -1 + \frac{1}{\sqrt{3}}, \, y_{1, -1} = -\frac{2}{3} \} \nonumber
\end{eqnarray}
and the same with $\beta \rar - \beta$, because we had started with the sign-reversal symmetric singular terms $u_{1,-1} \eqsp - u_{2,-1} \eqsp 1$ and so on. These solutions can painlessly be extended to higher orders in the $\beta$ expansion;  we have checked that they persist to $O(\beta^3)$ at least, with the finite/infinite rapidities showing an expansion in even/odd powers only. 

Taylor expanding the nested length-3 Bethe state for excitation numbers $\{4,3,2,2,1\}$ on these solutions, we see that due to the $m_5$ twist Lorentz invariance is broken in all cases beyond the leading terms at order $O(\beta^7)$. Worse still, the second and third solution are not even Lorentz invariant in the leading term. This is possibly a consequence of the co-existing $m_2$ twist which breaks $so(6)$, too. We do not consider these solutions any further; with hindsight we could have kept the condition $x_1 \eqsp - x_2$.

Fortunately, the first solution falls upon the desired result --- the leading order of the nested Bethe state becomes the Yukawa operator appearing in \eqref{ourL}! The customary normalisation \eqref{normalNorm} with the full Gaudin determinant ${\cal G}$ exactly sets off the $\beta^7$ of the Bethe vector and yields a real unit-norm state at the leading order. However, subleading orders of this norm are not useful. Normalising the state with cyclically identified kets by the coefficient of $|\bar Z \psi^{42} \psi^{31}\rangle$  instead we see a phase $e^{i \beta}$ for the Lorentz index pair 12 on the two fermions, and no such phase for 21. In conclusion, the twist regulator correctly computes the leading term but subleading orders do not relay any useful information. We would thus hope that any future hexagon computation works out at the leading order.

Can we reproduce the state from the matrix picture? An off-shell state of length 4 with the same excitation numbers can contain the following sets of fundamental excitations (for now no double occupation at any site)
\begin{eqnarray}
&& \{X, Y, \, \psi^{31}, \psi^{32}\} \, , \qquad \{\bar X, \bar Y, \, \psi^{41}, \psi^{42} \} \, , \nonumber \\
&& \{X, \bar X ,\psi^{31}, \psi^{42} \} \, , \qquad  \{X, \bar X, \psi^{32}, \psi^{41} \} \, , \\
&& \{Y,\, \bar Y,\, \psi^{31}, \psi^{42} \} \, , \qquad  \{Y, \, \bar Y, \, \psi^{32}, \psi^{41} \}  \nonumber
\end{eqnarray}
and their $S_4$ permutations. According to the usual matrix-picture recipe there will be one wave function for each initial sequence, and --- to be determined in some fashion --- a total of 144 coefficients  $g_{X Y \psi^{31} \psi^{32}}$ etc.\ multiplying these wave functions. On the other hand, appealing to cyclic invariance, there are only 24 distinct ket states in the physical $L \eqsp 3$ operator so that the problem of determining the $g$ coefficients by direct comparison is hugely underdetermined. 

Looking at the analytic off-shell results for the coefficients in primary states collected so far (see Section 1 and \cite{higherRank}) one would guess that the amplitudes are given by the coefficient of the first ket without double excitations in a nested Bethe wave of sufficient length, i.e.\ $| X Y \psi^{31} \psi^{32} Z^{L-4}\rangle$. This guess fails for the length-4 off-shell Bethe vector. It is a pressing open question what the general off-shell form of such coefficients is.

In the absence of a hypothesis for the exact form, we attempt to identify on shell like in Section \ref{potentiallySuper}, though now on the first Bethe solution in \eqref{all3}, \eqref{sixSolutions}, and we will do so only at leading order $\beta^0$. Without cyclic identification there are potentially 72 equations from the different kets in the length-3 state, but the constraints are quite degenerate. A central observation in \cite{higherRank} is that the $g$ coefficients are independent of the operator length, like the creation amplitudes and the $S$-matrices in the nested Bethe ansatz as well as in the matrix picture. We can therefore study the system of equations by substituting the Bethe solution and expanding in $\beta$ as stated. For higher-length states these may contain more information because of the higher number of kets; note also that they need not become cyclic on the solution. At length 4 this is still not a really difficult problem for {\tt Mathematica}, but at $L \eqsp 5$ it already becomes more involved. Limiting to ket states without double excitations --- recall that these carry all the information --- the length-5 problem is not too much larger than the full length-4 one: we count 720 differing kets now, as opposed to 408 before. Again, the equations are quite redundant and {\tt Mathematica} will choose certain coefficients to solve for. 

Labelling the coefficients of the 6 sets of 24 wave functions as $a_1 \ldots a_{24}, \, b_1 \ldots b_{24}, \ldots$ and putting free parameters to zero we can observe on the solution to the length-5 problem that the identifications
\beq
a_i \eqsp -b_i , \qquad c_i \eqsp - d_i \, \eqsp - e_i \eqsp f_i
\eeq
will hold. The conditions $d_i \eqsp - c_i, \, f_i \eqsp - e_i$ are expected from Lorentz invariance. Furthermore $b_i \eqsp - a_i$ and $e_i \eqsp -c_i$ are plausible from flipping $1 \, \leftrightarrow \, 2$ for the $su(2)$ indices of the $S$-matrix \cite{beisertSu22}: the minus sign stems from $X \rar -\bar X$ under this map.
Our toy example in Section 1, the Konishi singlet problem, also showed this last pattern, see equation \eqref{kon1Coef}. Finally, the sums over permutations of $\{X, Y, \, \psi^{31}, \psi^{32}\}, \, \{\bar X, \bar Y, \, \psi^{41}, \psi^{42} \}$, respectively, should separately be Lorentz invariant, reducing the number of independent $a_i$ coefficients to 12 and thus we have shrunk the problem to determining the more feasible number of 36 independent unknown coefficients.

The solution is (allowing for distinct normalisations at $L \eqsp 3,4,5$): 
\begin{eqnarray}
0 & = & \phm a_{X Y \psi^{31} \psi^{32}} + a_{Y X \psi^{31} \psi^{32}} \eqsp \phm a_{X Y \psi^{32} \psi^{31}} + a_{Y X \psi^{32} \psi^{31}} \, , \nonumber \\
2 & = & -a_{X \psi^{31} Y \psi^{32}} -a_{\psi^{31} X Y \psi^{32}} \eqsp -a_{Y \psi^{32} X \psi^{31}} -a_{\psi^{32} Y X \psi^{31}}  \, , \nonumber \\
2 & = & \phm a_{X \psi^{31} \psi^{32} Y} + a_{\psi^{31} X \psi^{32} Y} \eqsp \phm a_{Y \psi^{32} \psi^{31} X} + a_{\psi^{32} Y \psi^{31} X}  \, , \label{longPhPsPsA} \\
2& = & \phm a_{Y \psi^{31} X \psi^{32}} + a_{\psi^{31} Y X \psi^{32}} \eqsp \phm a_{X \psi^{32} Y \psi^{31}} + a_{\psi^{32} X Y \psi^{31}}  \, , \nonumber \\
2 & = & -a_{Y \psi^{31} \psi^{32} X} -a_{\psi^{31} Y \psi^{32} X} \eqsp -a_{X \psi^{32} \psi^{31} Y}-a_{\psi^{32} X \psi^{31} Y}  \, , \nonumber \\
16 & = & -a_{\psi^{31} \psi^{32} X Y} + a_{\psi^{32} \psi^{31} X Y} \eqsp \phm a_{\psi^{31} \psi^{32} Y X} -a_{\psi^{32} \psi^{31} Y X} \, , \nonumber
\end{eqnarray}
and
\begin{eqnarray}
0 & = & \phm c_{X \bar X \psi^{31} \psi^{42}} +  c_{\bar X X \psi^{31} \psi^{42}} \eqsp \phm c_{X \bar X \psi^{42} \psi^{31}} + c_{\bar X X \psi^{42} \psi^{31}}  \, , \nonumber \\
2 & = & -c_{X \psi^{31} \bar X \psi^{42}} -c_{\psi^{31} X \bar X \psi^{42}} \eqsp -c_{\bar X \psi^{42} X \psi^{31}} -c_{\psi^{42} \bar X X \psi^{31}} \, , \nonumber \\
2 & = & \phm c_{X \psi^{31} \psi^{42} \bar X}  + c_{\psi^{31} X \psi^{42} \bar X}  \eqsp \phm c_{\bar X \psi^{42} \psi^{31} X} + c_{\psi^{42} \bar X \psi^{31} X} \, , \label{longPhPsPsB} \\
4 & = & -c_{X \psi^{42} \bar X \psi^{31}} -c_{\psi^{42} X \bar X \psi^{31}} \eqsp  -c_{\bar X \psi^{31} X \psi^{42}} -c_{\psi^{31} \bar X X \psi^{42}} \, , \nonumber \\
4 & = & \phm c_{X \psi^{42} \psi^{31} \bar X} + c_{\psi^{42} X \psi^{31} \bar X} \eqsp \phm c_{\psi^{31} \bar X \psi^{42} X} + c_{\bar X \psi^{31} \psi^{42} X} \, , \nonumber \\
8 & = & \phm c_{\psi^{31} \psi^{42} X \bar X} -c_{\psi^{42} \psi^{31} X \bar X} \eqsp -c_{\psi^{31} \psi^{42} \bar X X} + c_{\psi^{42} \psi^{31} \bar X X} \, .  \nonumber
\end{eqnarray}
Reality and unit norm of the state are achieved by an overall factor $i/(432 \sqrt{2})$. The apparent 24-parameter freedom does not change the state because the wave functions are degenerate: 
\beq
\Psi_{ABCD} \eqsp \Psi_{BACD}
\eeq
as long as one of $A,B$ is bosonic and
\beq
\Psi_{ABCD} \eqsp - \Psi_{BACD}
\eeq
for two fermions $A,B$. This happens because the first two excitations come with the infinite rapidities $u_1,u_2$. At leading twist they behave like free particles that (anti-) commute. To economise on the workload in future applications we can safely put one half of the coefficients in \eqref{longPhPsPsA}, \eqref{longPhPsPsB} to zero.

\subsection{The Yang-Mills term} \label{YMconjecture}

We aim to recover the linear combination (written out without appealing to cyclic symmetry)
\beq
|F^{11} F^{22} \ra - 2 |F^{12} F^{12} \ra + |F^{22} F^{11}\ra \label{dreamF2}
\eeq
from the nested as well as the matrix Bethe ansatz, cf.\ Section \ref{secL}. In the nested Bethe ansatz corresponding to Figure 1, panel 3, we need four level-1 rapidities and two auxilliary rapidities on either wing to reach $F$. In constructing the nested Bethe ansatz, we generate the same elements as before: the level-1 rapidity moves over the chain with the momentum factor \eqref{momFac}, but since the excitation is fermionic we have $S^{11} \, = \, -1$ (i.e.\ the well-known $su(1|1)$ problem). Choosing $Q_3^2$ as a lowering operator (Lorentz index $\alpha \eqsp 2$) the level-1 excitation is $\psi^{42}$. On the left wing, Lorentz generators $L_2^1$ move, while on the right wing there is the $so(6)$ generator $R_4^3$, and these operators commute. On the left wing we have creation amplitudes and $S$-matrices as for the Lorentz generator in the fermion problem of the last section, thus as in the second line of \eqref{level45bits}. On the right wing the definitions are as for the $so(6)$ generators in the introduction. 

(Multiple) excitations can create the following fields:
\vskip 0.2 cm
\begin{center}
\begin{tabular}{c|c|l} 
field & $\{n_1,n_2,n_3\}$ & extra amplitude \\
\hline
$\psi^{42}$ & $\{1,0,0\}$ & $\phm 1$ \\[1 mm]
$\psi^{41}$ & $\{1,1,0\}$ & $\phm 1$ \\[1 mm]
$\psi^{32}$ & $\{1,0,1\}$ & $\phm 1$ \\[1 mm]
$\psi^{31}$ & $\{1,1,1\}$ & $\phm 1$ \\[1 mm]
$F^{22}$ & $\{2,0,1\}$ & $ p_1 \eqsp -(u_1-u_2)$ \\[1 mm]
$F^{12}$ & $\{2,1,1\}$ & $ p_2 \eqsp \phm (u_1-u_2)(u_1+u_2-2 \, v_1)$ \\[1 mm]
$F^{22}$ & $\{2,2,1\}$ & $ p_3 \eqsp \phm (u_1 - u_2) (v_1-v_2) \bigl([v_1-v_2]^2 - 4 \, u_1 u_2 + 1\bigr) / \bigl(2 \, [v_1-v_2 - i]\bigr)$
\end{tabular}
\end{center}
As before we have estimated the polynomial order by comparing terms with double excitations to other kets, in which the level-1 magnons do not share the same site. This estimate makes us expect extra numerator polynomials of linear, quadratic and cubic order, respectively, for $F^{22}, F^{21}, F^{11}$. Using states with excitation numbers $\{2,0,1\}$, it is easy to interpolate the linear polynomial for the creation of $F^{22}$.  The next higher primary states seem to appear only at excitation numbers $\{4,2,2\}$ as for the state we have in mind, but for $L \, \geq 4$. We have used states at lengths 4,5,6 with rapidity pairs only to determine the remaining two numerators by interpolation. While the amplitude for $F^{12}$ is not hard to obtain, one comes to realise that $p_3$ for $F^{22}$ is not polynomial. The next best guess turns out to work, namely fourth order over a linear denominator. To simplify the task we split off two linear factors as in the interpolation for $q$, see \eqref{oriQ}. Somewhat surprisingly, $p_2$ is not a part of $p_3$, whereas the choice
$(u_1-u_2)(v_1-v_2)$ is apparently correct. Due to the antisymmetric choice $v_1 \eqsp -v_2$ this interpolation can in fact only be unique employing a single variable $2 \, v \eqsp v_1 - v_2$ yielding
\beq
p_3 \eqsp \frac{(u_1 - u_2) \, v \, (4 \, v^2 - 4 \, u_1 u_2 + 1)}{2 \, v - i} \, .
\eeq
The result in the table is its natural generalisation.

Next, we need a solution of the $L \eqsp 2$ Bethe equations for $\{4,2,2\}$ magnons. Our operator is a vacuum descendent whence \emph{all} rapidities ought to be infinite. As for the Yukawa term, in order to avoid equal rapidities (which would annihilate the Bethe state) we introduce twists $m_0 \ldots m_3$ on the momentum node and the three nodes of the Dynkin diagram. Once again we try to simplify the equations by looking for pairs $\pm u$ etc. We could not find solutions with rapidity pairs only, the deviating pattern for the $x$'s in  \eqref{sixSolutions} now occurs in the level-3 rapidities $w_j$. For real twists \beq
\{ m_0 = 0, \, m_2 = 0, \ u_{1,-1} = \frac{1-i}{\sqrt{2}}, \, u_{2,-1} = \frac{1+i}{\sqrt{2}}, \, u_{3,-1} = - \frac{1+i}{\sqrt{2}}, \, u_{4,-1} = - \frac{1-i}{\sqrt{2}} \} \label{solF2a}
\eeq
and 
\begin{eqnarray}
&& \{ m_1 = -\frac{\sqrt{2}}{t}, \, m_3 =  \phm \frac{2 \sqrt{2}}{t}, \ v_{1,-1} = -t, \, v_{2,-1} = t, \,
      w_{1,-1} = -t \, (\sqrt{2} + i), \, w_{2,-1} = -t \, (\sqrt{2} - i) \} \, , \label{solF2b} \\
&& \{ m_1 =  \phm \frac{\sqrt{2}}{t}, \, m_3 = -\frac{2 \sqrt{2}}{t}, \ v_{1,-1} = -t, \, v_{2,-1} = t, \,
      w_{1,-1} =  \phm t \, (\sqrt{2} - i), \, w_{2,-1} = \phm t \, (\sqrt{2} + i) \} \nonumber
\end{eqnarray}
with $t \, = \, 3^{-1/4}$. These solutions, too, can be extended at least through $O(\beta^3)$. There will be non-trivial corrections to the rapidities only in odd powers of $\beta$. There are similar solutions with imaginary $m_j, \, v_j$ and a factor $i$ on the $w_j$, too. Taking a closer look we see that these are related by sending $\beta \rar \beta \, i^n$, which is the $\mathbb Z_4$ invariance of the set $\{ u_{j,-1} \}$. Since there is no preferred scale for $\beta$ the four solutions are indeed equivalent.
Expanding the Bethe state on the first solution in \eqref{solF2a}, \eqref{solF2b} and normalising by \eqref{normalNorm} (both, state and norm are of order $O(\beta^4)$) we obtain the pure Yang-Mills Lagrangian \eqref{dreamF2} with normalisation $1/\sqrt{24}$.

To achieve the same in the matrix picture we have to construct creation amplitudes for the double excitations $F^{\alpha \beta}$. The states we used for the purpose of interpolation in the nested picture are perfectly suitable here as well. Compatibility is achieved by
\beq
- e_{\psi^{3\alpha} \psi^{4\beta}} \eqsp e_{\psi^{4\alpha} \psi^{3\beta}} \eqsp U_{12} \, .
\eeq
The 24 possible amplitudes start out the $S_4$ permutations of the magnons $\{\psi^{31} \psi^{32} \psi^{41} \psi^{42}\}$ with fixed initial rapidity ordering $\{u_1,u_2,u_3,u_4\}$. However, all rapidities are infinite and thus at leading twist all particles are free. As a consequence, the 24 wave functions are equal up to signs given by the signature of the permutations. We can reproduce the pure Yang-Mills Lagrangian choosing any one amplitude! The $g$ coefficient will equal $\pm 1/\sqrt{24}$ to match the normalised nested ansatz or $\pm \beta^4$ when the standard normalisation \eqref{normalNorm} is dropped.

\section{A first test of $\Psi \bigl(\psi^{42} \psi^{41} \psi^{32} \psi^{31}\bigr)$ in the r\^ole of the Lagrangian} \label{amazingTest}

Absorbing a factor of $g$ into all fields in the \nFour super Yang-Mills Lagrangian \eqref{ourL}, we can write the path integral for an $n$-point function of gauge-invariant composite operators as
\beq
\la \, \cO_1 \ldots \cO_n \, \ra \eqsp \int D \phi \, D A \, D \psi  \ e^{\frac{i}{g^2} \int d^4x_0 {\cal L}(x_0)} \ \cO_1 \ldots \cO_n \, .
\eeq
It follows that
\beq
\frac{\partial}{\partial g^2} \la \, \cO_1 \ldots \cO_n \, \ra \eqsp - \frac{i}{g^4} \int d^4x_0 \, \la \, {\cal L}_0 \; \cO_1 \ldots \cO_n \, \ra~. 
\eeq
Restricting to leading order, this implies that the one-loop correction to an $n$-point function can be computed from the Born level of an $n+1$-point function in which an integrated Lagrange operator is inserted. That the right-hand side is indeed a tree-level contribution (lowest possible order) is a consequence of the field rescaling that causes a factor $g^2$ on each propagator. One can undo the field rescaling without hampering the functional relation between the two correlation functions.

Two-point functions of BPS operators are protected, see \cite{nilpotent} and the references therein. Hence it must be so that
\beq
\la \, {\cal L}_0 \; \cO^L_1 \, \cO^L_2 \, \ra \eqsp 0 \label{LOO}
\eeq
for two spin-chain vacua $\cO^L$. In the following we will use this identity as a non-trivial test of our approach, and indeed will find agreement! 


In the hexagon computation we must first cut a Bethe state with the four magnons $\{\psi^{42}, \psi^{41}, \psi^{32}, \psi^{31} \}$ into two parts. Apart from the momentum shift $e^{i \, p \, l}$ that every magnon picks up jumping from the first stretch to the second, the form of the entangled state is length-independent and there is no reference to double excitations. We can thus study a length-4 (or higher) spin chain with four elementary excitations moving on it. Nevertheless, the scattering between the magnons is quite complicated because the $F_{12}$ and $ C_{12}$ elements of the $S$-matrix can transform a pair of bosons into a pair of fermions and back, respectively. As there are no magnons from the two other operators in the three-point function, the two sets in a partition $(\alpha, \bar \alpha)$ of the excitations are then each submitted to one hexagon operator.

The operator $\tr(F^2)$ is chiral: there are four fermions on the left chain and four bosons on the right chain. Therefore, any non-vanishing contribution to the three-point function \eqref{LOO} must be carried by at least two $C_{12}$ or $F_{12}$ processes and is therefore of order $g^2$. If 0/1/2 of these arise from the scattering of the magnons resulting from the construction of the entangled state, then 2/1/0 $F,C$'s must be contributed by the scattering on the associated product of two hexagons. We can thus work to leading order in the Yang-Mills coupling $g$ and the twist $\beta$ in \emph{all} terms. The effect is a drastic reduction of complexity: the only contributing partitions are 
\begin{eqnarray}
\la \, {\cal L}_0 \; \cO^L_1 \, \cO^L_2 \, \ra &= & 2 \, \Bigl[ \la  \mathfrak{h} \, | \, \psi^{42}_1 \, \psi^{41}_2 \, \psi^{32}_3 \, \psi^{31}_4 \, \ra + \la \mathfrak{h} \, | \, \psi^{42}_1 \, \psi^{31}_4 \, \ra \ \la \mathfrak{h} \, | \, \psi^{41}_2 \, \psi^{32}_3 \, \ra \, \Bigr] +  \\
&& \tilde g \, \Bigl[  \la \mathfrak{h} \, | \, D^{2 \dot 1}_1 \, \ra \  \la \mathfrak{h} \, | \, \psi^{41}_2 \, \psi^{32}_3 \, D^{1 \dot 2}_4 \, \ra + \la \mathfrak{h} \, | \, D^{1 \dot 2}_2 \, \ra \  \la \mathfrak{h} \, | \, \psi^{42}_1 \, D^{2 \dot 1}_3 \, \psi^{31}_4 \, \ra +\nonumber \\
&& \phantom{2 \, \Bigl[} \la \mathfrak{h} \, | \, D^{2 \dot 1}_3 \, \ra \  \la \mathfrak{h} \, | \, \psi^{42}_1 \, D^{1 \dot 2}_2 \, \psi^{31}_4 \, \ra + 
\la \mathfrak{h} \, | \, D^{1 \dot 2}_4 \, \ra \  \la \mathfrak{h} \, | \, D^{2 \dot 1}_1 \,  \psi^{41}_2 \, \psi^{32}_3 \, \ra +
\nonumber \\
 &&  \phantom{2 \, \Bigl[} \la \mathfrak{h} \, | \,Y_1 \, \ra \  \la \mathfrak{h} \, | \, \psi^{41}_2 \, \psi^{32}_3 \ \bar Y_4 \, \ra +
 \la \mathfrak{h} \, | \,Y_2 \, \ra \  \la \mathfrak{h} \, | \, \psi^{42}_1 \,  \bar Y_3 \,  \psi^{31}_4 \, \ra \; +  \nonumber \\
&& \phantom{2 \, \Bigl[} \la \mathfrak{h} \, | \,\bar Y_3 \, \ra \ \la \mathfrak{h} \, | \,\psi^{42}_1 \,  Y_2 \,  \psi^{31}_4 \, \ra + 
\la \mathfrak{h} \, | \,\bar Y_4 \, \ra \ \la \mathfrak{h} \, | \,Y_1 \,  \psi^{41}_2 \,  \psi^{32}_3 \, \ra \Bigr] + \nonumber \\
&& \tilde g^2 \Bigl[ \la \mathfrak{h} \, | \,D^{2 \dot 1}_1 \,  D^{1 \dot 2}_2 \, \ra \ \la \mathfrak{h} \, | \,D^{2 \dot 1}_3 \,  D^{1 \dot 2}_4 \, \ra +  \la \mathfrak{h} \, | \,D^{2 \dot 1}_1 \,  Y_2 \, \ra \ \la \mathfrak{h} \, | \,\bar Y_3 \,  D^{1 \dot 2}_4 \, \ra + \nonumber \\
&& \phantom{\tilde g^2 \Bigl[} \la \mathfrak{h} \, | \,Y_1 \,  D^{1 \dot 2}_2 \, \ra \ \la \mathfrak{h} \, | \,D^{2 \dot 1}_3 \,  \bar Y_4 \, \ra  +  \la \mathfrak{h} \, | \,Y_1 \,  Y_2 \, \ra \ \la \mathfrak{h} \, | \,\bar Y_3 \,  \bar Y_4 \, \ra \, \Bigr]~. \nonumber 
\end{eqnarray}
The effective coupling $\tilde g \eqsp \sqrt{g^2 N} \, i \, \beta^2/\sqrt{2}$ arises from the normalisation of the $C_{12}, F_{12}$ elements and their unbalanced denominator factors $x_1^-  x_2^-, \, x_1^+ x_2^+$, respectively. Inserting the hexagon amplitudes, the three parts go to
\beq
\la \, {\cal L}_0 \; \cO^L_1 \, \cO^L_2 \, \ra \rar 4 \, \tilde g^2 \, (1 - 2 + 1) \eqsp 0 \label{doesWork}
\eeq
as desired. Curiously, the sign of the individual parts does not flip according to the signature of the permutation when we choose a different initial flavour ordering of the magnons. This already implies that their sum must vanish.

We conclude the section with a couple of comments about the features of the computation. First, the overall order $g^2$ signals a one-loop computation, so we are on the right track. In an upcoming study we will study the insertion $\tr(F^2)$ into a two-point function of scalar two-excitation BMN operators \cite{bmn} in order to provide further tests of our approach. With respect to the order counting that computation should be very similar to what happened here. This clearly indicates that the Bethe state representing $\tr(F^2)$ does incorporate the admixtures (Yukawa and superpotential) in the field-theory result \eqref{ourL}. Second, the normalisation fits the usual pattern: for the non-normalised wave function we have an overall coefficient $g_{\psi^{42} \psi^{41} \psi^{32} \psi^{31}} \, \sim \, \beta^4$, so the total order is $O(\beta^8)$ which will be offset by the root of the full Gaudin determinant when normalising the hexagon amplitude. Alternatively, we can normalise the nested Bethe state to 1, upon which $g_{\psi^{42} \psi^{41} \psi^{32} \psi^{31}} \eqsp 1/\sqrt{24}$ and thus the computation stays at $O(\beta^4)$ as visible in \eqref{doesWork}. This is compensated by Korepin's normalisation factor $\sqrt{\prod (u_i^2 + 1/4)} \rar 1/\beta^4$ in accordance with equation \eqref{agree}. Third, we have evaluated the hexagons in the spin-chain frame paying due attention to $Z$ markers \cite{beisertSu22,BKV}. On the other hand, we were lenient about the $Z$ markers in deriving the entangled state. This is justified in the case at hand because all momentum factors go to 1 in the limit $\beta \rar 0$. In fact, the edge width as a quantum number in the entangled state drops out. When there are magnons with non-zero momenta the consequences of length changing will become a subtle issue. 


\section{Conclusions}

When there are several flavours of excitations on a spin chain, the scattering involves a true $S$-\emph{matrix} of phases arising when one magnon passes another. In this picture one writes a multi-component wave function for all possible initial ordering of flavours, with the same initial orderings of rapidities for all parts of the ansatz. Each partial wave function comes with a normalisation coefficient called $g$ in the above. In the process of solving the associated matrix Bethe equations, these will eventually be fixed. The nested Bethe ansatz is a method for diagonalising the scattering, where a single rather involved wave function is introduced. The two pictures must agree but are not manifestly equivalent. We can extract the $g$ coefficients of the matrix picture upon equating with the nested ansatz.

The hexagon formalism for structure constants of \nFour super Yang-Mills theory is most naturally written in the matrix picture. Importing the $g$ coefficients determined by comparing operators as outlined above, it becomes possible to exploit it for operators in higher-rank sectors, i.e.\ featuring magnons of various flavours \cite{higherRank}. 

In the integrable system for the planar spectrum of the AdS(5)/CFT(4) duality \cite{beiStau}, a scalar, say $Z$, of the \nFour super Yang-Mills theory is chosen as a spin-chain vacuum. This breaks the symmetry algebra $psu(2,2|4)$ to a $psu(2|2)_L \otimes psu(2|2)_R$ subalgebra. Importantly, in this process the conjugate $\bar Z$ of the vacuum scalar, as well as half the fermionic excitations, and the chiral and anti-chiral field strength tensors $F, \tilde F$ are lost as possible excitations in the spin-chain problem. In this article we have argued that they can be recovered as double excitations in either picture, the nested Bethe ansatz as well as the matrix version. In the nested ansatz this is possible at least at tree level, allowing all three simple roots $R_2^1, R_3^2, R_4^3$ of the internal $SO(6) \, \cong \, SU(4)$ symmetry to occur as double excitations. The inclusion of higher-loop effects is possible only for certain gradings of the symmetry algebra, for which we would have to review the way the Bethe ansätze are organised in the present work. On the other hand, the matrix picture is much more agnostic so that completing the universal creation amplitude $U_{12}$ in \eqref{extraF} to an expression in terms of the coupling-dependent Zhukowky variables $x^\pm$ \cite{beiStau} is very likely possible. We should study local terms also in Bethe wave functions with multiple derivative excitations, a question that was excluded here. An interesting aside is the question whether the extra amplitudes in the matrix picture can help to lift the restriction to only four possible gradings for the construction of the all-loops asymptotic nested ansatz in \cite{beiStau}.

As an application of our ideas, we have expressed the three types of operators appearing in the mixing problem of the \nFour on-shell Lagrange density as Bethe states. The main focus was on the chiral Yang-Mills Lagrangian $\tr(F^2)$. The effects of its field-theory admixtures \eqref{ourL} --- a Yukawa term and the scalar superpotential --- should in fact be incorporated in coupling corrections to elements of the Bethe ansätze and the relevant solution. In future work we hope to reach clarity on this issue.

Meanwhile our main result is that, owing to the existence of double occupations, the chiral field strength $\tr(F^2)$ can be represented at least at tree level as a length-2 state with the four fermionic excitations
\beq
\{\psi^{42}, \psi^{41}, \psi^{32}, \psi^{31} \} \, ,
\eeq
all with infinite level 1-rapidities as stated in \eqref{solF2a}. In both Bethe pictures the particles are then free. As a consequence, the 24 possible wave functions (differing by the flavour ordering) become fully degenerate, so that we can select any one of them. Second, in the matrix picture not only the (diagonal part of the) scattering matrix but also our new creation amplitudes \eqref{extraF} scale to 1 if only one of the rapidities is taken to infinity. Therefore even the normalisation of the one wave function we choose can easily be determined comparing to any one data point.

What is more, the hexagon formalism can be fed with the information stated without mentioning the double excitations at all, cf.\ \cite{higherRank} or the example discussed in Section 1 of this work. We are confident that three- or indeed higher-point functions with insertions of the Lagrange operator can be addressed by these methods. Excitingly, this opens a window to the inclusion of non-planar corrections into the spectrum problem by Lagrangian insertion on a tiling \cite{cushions, shotaThiago1,colourDressed,handles1} or in the ideal case even to the simplification of gluing corrections \cite{BKV,fivePoints}. Our conjecture for $\tr(F^2)$ passes a first test \eqref{doesWork} presented in Section \ref{amazingTest}: half-BPS two-point functions are protected, whereby the three-point function with two equal vacua and a Lagrangian insertion must vanish.

We leave more elaborate computations to future work. Here possible further complications, e.g.\ length-changing effects, how to handle the twist regulator we used for the Bethe solutions in hexagon computations \cite{doubleTorus,higherRank}, the proper inclusion of loop effects, or simply establishing comparable field-theory results.

Last, the construction presented in \cite{higherRank} and extended in the present article is interesting in its own right. One important question is to determine a sufficiently general analytic form for the relative coefficients of the component wave functions in the matrix picture, hence a suitable generalisation of \eqref{kon1ana}.

\section*{Acknowledgements}

We are grateful to T.~McLoughlin, A.~Sfondrini and M.~Staudacher for discussions clarifying some aspects of the material here presented. B.~Eden is supported by Heisenberg funding of the Deutsche Forschungsgemeinschaft, grant Ed 78/7-1 or 441791296. D.~le~Plat is supported by the Stiftung der Deutschen Wirtschaft. A.~Spiering is supported by the research grant  00025445 from Villum Fonden, as well as by the European Union's Horizon 2020 research and innovation program under the Marie Sk\l odowska-Curie grant agreement No.\ 847523 `INTERACTIONS'.

\end{document}